\documentclass[pre,a4paper,twocolumn,notitlepage,nofootinbib]{revtex4-1}
\usepackage{amsmath}
\usepackage{amsfonts}
\usepackage{bbm}
\usepackage{bm}
\usepackage{latexsym}
\usepackage{graphicx}
\usepackage{subfigure}

\newcommand{\uni}[1]{\,\mathrm{#1}}
\newcommand{\beq}{\begin{equation}}
\newcommand{\eeq}{\end{equation}}

\begin{document}

\title{\bf Maximal-entropy random walk unifies centrality measures}

\author{J.K. Ochab}\email{jeremi.ochab@uj.edu.pl}


\affiliation{Marian Smoluchowski Institute of Physics\\
Jagiellonian University, Reymonta 4, 30-059 Krak\'ow, Poland}

\date{\today}

\begin{abstract}
In this paper analogies between different (dis)similarity matrices are derived.
These matrices, which are connected to path enumeration and random walks,
are used in community detection methods or in computation of centrality measures for complex networks.
The focus is on a number of known centrality measures, which inherit the connections established for similarity matrices.
These measures are based on the principal eigenvector of the adjacency matrix, path enumeration,
as well as on the stationary state, stochastic matrix or mean first-passage times of a random walk.
Particular attention is paid to the maximal-entropy random walk,
which serves as a very distinct alternative to the ordinary random walk used in network analysis. 

The various importance measures, defined both with the use of ordinary random walk and the maximal-entropy random walk,
are compared numerically on a set of benchmark graphs.
It is shown that groups of centrality measures defined with the two random walks
cluster into two separate families.
In particular, the group of centralities for the maximal-entropy random walk,
connected to the eigenvector centrality and path enumeration,
is strongly distinct from all the other measures and produces largely equivalent results.

\medskip


\end{abstract}

\maketitle
		
\section{Introduction}

Graphs represent abstracted relationships between entities.
They form a structure upon which a process may take place,
which is often formalized into the mathematical concept called random walk.
Together, graph and random walk can constitute a model
for citations in scientific collaboration networks, opinion spread on social networks or data transmission on the Internet.
Instead of these kinds of information transfer, more tangible subjects may be considered,
as molecule movement on physical or biological networks.
Whatever the exact nature of the phenomenon, the natural question arises:
which entity in the network is the most influential,
be it a gene or a transcription factor, an overloaded hub, a frequented website or a renowned researcher.

A number of importance (or centrality) measures answering that question
have been invented to study social (e.g., \cite{Freeman}; \cite{WF} is an extensive resource)
or telecommunication networks (e.g., HITS \cite{HITS}, PageRank \cite{Page}).
A significant portion of ideas defining the measures originate from graph theory
(the degree of a vertex, enumeration of paths or the principal eigenvector of the adjacency matrix)
and the theory of Markov chains (stationary states of random walks, their stochastic matrices, and mean first-passage times).
Likewise, most of these approaches have been widely utilized in algorithms of community finding \cite{Fortunato}.

In this note, we show that a number of these ideas can be formulated in a common framework,
they produce nearly equivalent results for a given random walk,
and additionally, the results are more distinct from other methods
if they make use of the maximal entropy random walk (MERW).

This random walk (RW) is defined so as to ensure equiprobability of all paths of a given length and endpoints.
Other RWs usually do not demand as much,
for instance equal probabilities of going from a node to any of its nearest neighbors are enough
to define what is called here the generic random walk (GRW),
one that is well-known and commonly used.
It is often the case that either GRW or some biased RWs are better suited for particular problems.
However, the author believes MERW should serve alongside GRW as a null model of random processes on networks,
for a good reason:
it is GRW that maximizes entropy locally (entropy of the nearest neighbor selection),
and it is MERW that maximizes entropy globally (entropy of the path selection).
Since a random walk can be seen as an ensemble of possible paths,
it is the latter that yields the largest entropy for that ensemble \cite{ZB1,ZB2}.
Thus, in a sense, it is the most random of random walks.

MERW exhibits behaviors that may be of general interest: its stationary distribution localizes on diluted lattices \cite{ZB1, ZB2, BW},
its relaxation to stationary state is extremely fast on Cayley trees \cite{ZB3,Demo2},
and it is very slow between two identical connected k-regular regions \cite{JO}.
The equiprobable paths that MERW produces are the natural candidates for an ensemble used in Feynman path integrals (in models of discrete quantum gravity with curved space-time) \cite{ZB2} or in the optimal sampling algorithm in the path-integral Monte Carlo methods \cite{H}.
Since entropy maximization is a global principle, conceptually analogical to the least action principle, it was also studied in biology and has led to the concept of evolutionary entropy \cite{Evolutionary1}.
The same authors have found the value of entropy for a given graph useful in selection of robust networks \cite{Evolutionary2}.
Lastly, MERW has begun to be used as a tool for analysis of complex networks \cite{MERW+CN1, Delvenne, MERW+CN3, MERW+CN4, MERW+CN5}.

\section{Generic and maximal-entropy random walks}
Let us consider a finite connected undirected graph.
We define a discrete-time random walk on this graph by a stochastic (or transition) matrix $\mathbf{P}$.
Its entry $P_{ij} \ge 0$ is the probability
that if a random walker stays on a node $i$ at a time $t$,
it will step to a node $j$ at time $t+1$.
Any row $P_{i*}$ contains the probabilities of moving to all neighbors of $i$,
and since the walker cannot disappear from the graph nor be created,
they all sum up to unity, $\sum_j P_{ij} = 1$.
As we assume that the walker can only move to neighboring nodes,
the stochastic matrix can have a non-zero entry only if the adjacency matrix of the graph has a non-zero entry at the same place.
Shortly, $\forall i,j: P_{ij}\leq A_{ij}$, where $\mathbf{A}$ is the adjacency matrix of the graph.
Elements of this matrix can take two values: $A_{ij}=1$ if $i$ and $j$ are neighbors and $A_{ij}=0$ otherwise.
Both $\mathbf{P}$ and $\mathbf{A}$ are assumed to be time-independent.

The probability that the random walker stays at a given vertex $i$ of the graph at a given time $t$
is encoded in the $i$-th element of the vector $\vec{\mathbf{\pi}}(t)^T=(\pi_1(t),\ldots,\pi_N(t))$.
Thus, the initial distribution of probabilities is $\vec{\mathbf{\pi}}(0)^T$,
and the distribution after $t$ steps $\vec{\mathbf{\pi}}(t)^T=\vec{\mathbf{\pi}}(t-1)^T\mathbf{P}=\vec{\mathbf{\pi}}(0)^T\mathbf{P}^t$,
where the stochastic matrix has been multiplied $t$ times.

A quantity of interest, given by a solution of
\beq
\vec{\pi}^T=\vec{\pi}^T\mathbf{P} \ ,
\eeq
is the stationary probability distribution (or stationary state),
which may be understood as the probability distribution after infinite time.
We assume it exists
\footnote{A stationary state exists if an undirected graph is not bipartite,
but even for bipartite graphs a semi-stationary state can be defined
by averaging probability distribution over two consecutive time steps.}.

The ordinary or, as we call it, generic random walk
corresponds to the standard random walks used by Einstein, Smoluchowski or Polya.
It is realized by the following stochastic matrix
\beq
P_{ij} = \frac{A_{ij}}{k_i} \ ,
\label{eq:Porw}
\eeq
where $k_i = \sum_j A_{ij}$ denotes a degree of $i$-th node.
Its stationary state is proportional to the degrees and is given by $\pi_i = k_i/\sum_j k_j$.
An $i$-th row of the matrix contains uniform probabilities, each equal to $1/k_i$,
of selecting any of the $k_i$ neighbors of the node $i$.
Thus, the entropy of neighbor selection is maximal.

The other type of RW introduced earlier, maximizes the entropy of random trajectories,
and hence is called here the maximal-entropy random walk.
This maximization condition leads to a unique stochastic matrix
\beq
\label{eq:Pmerw}
P_{ij} = \frac{A_{ij}}{\lambda_0} \frac{\psi_{0j}}{\psi_{0i}} \ ,
\eeq
where $\lambda_0$ is the largest eigenvalue of the adjacency matrix $\mathbf{A}$,
and $\psi_{0i}$ is the $i$-th element of the principal eigenvector $\vec{\psi}_0$.
Since the adjacency matrix is irreducible,
the Frobenius-Perron theorem guarantees that all elements of this vector are strictly positive,
thus the condition $P_{ij}\leq A_{ij}$ is fulfilled.

MERW has the stationary probability distribution given by Shannon-Parry measure \cite{P}
\beq
\label{eq:stat}
\pi_i = \psi^2_{0i} \ .
\eeq
Let us note that this formula allows to interpret $\psi_{0i}$ as the wave function of the ground state
of the operator $-\mathbf{A}$ and $\psi^2_{0i}$ as the probability of finding a particle in this state \cite{ZB1,ZB2},
thus relating MERW to quantum mechanics.

It is easily seen that the two RWs, \eqref{eq:Porw} and \eqref{eq:Pmerw},
are identical on $k$-regular graphs.
This should be considered an exception,
as in general their properties are entirely distinct and contrasting.


\section{Relations between the stochastic matrix, its distance matrix, mean first-passage time matrix, and the resolvent of adjacency matrix}
\label{sec:theory}

In general, a stochastic matrix may be not symmetric, and so it may have different right and left eigenvectors:
\beq
\mathbf{P}\vec{\Psi}_\alpha = \Lambda_\alpha \vec{\Psi}_\alpha \quad , 
\quad
\vec{\Phi}_\alpha^T \mathbf{P} = \Lambda_\alpha \vec{\Phi}_\alpha^T \ ,
\eeq
which results in a spectral decomposition 
\beq
\label{eq:Pspec}
\mathbf{P}=\sum_{\alpha}\Lambda_{\alpha} \vec{\Psi}_{\alpha}\vec{\Phi}_{\alpha}^T \ .
\eeq

Let us consider a class of Markov processes whose stochastic matrix can be transformed into a symmetric matrix
\beq
\label{eq:S}
\mathbf{S}=\mathbf{D}^{-1/2}\mathbf{P}\mathbf{D}^{1/2} \ ,
\eeq
where we define $\mathbf{D}\equiv \uni{diag}(\vec{\pi})^{-1}$,
which denotes the matrix with diagonal entries equal to the inverses of the stationary state vector's elements.
It follows that
\beq
\label{eq:PhiPsi}
\vec{\Phi}_{\alpha}=\mathbf{D}^{-1} \vec{\Psi}_{\alpha} \ .
\eeq
This relation does not hold in general but is clearly obtained for GRW and MERW,
which satisfy \eqref{eq:S}, as shown below.

For GRW, stochastic matrix \eqref{eq:Porw} can be written as
\beq
\mathbf{P}=\uni{diag} (k_1,k_2,\ldots,k_N)^{-1}\mathbf{A} \ ,
\eeq
where the adjacency matrix can be decomposed into $\mathbf{A}=\sum_{\alpha}\lambda_{\alpha} \vec{\psi}_{\alpha}\vec{\psi_{\alpha}}^T$.
Substitution of the stationary state of GRW into the diagonal matrix yields
$\mathbf{D}=\uni{diag} (k_1,k_2,\ldots,k_N)^{-1} \sum_{j}k_j \ ,$
hence
\beq
\mathbf{P}
=\frac{1}{\sum_{j}k_j}\sum_{\alpha}\lambda_{\alpha} \mathbf{D}\vec{\psi}_{\alpha}\vec{\psi_{\alpha}}^T \ ,
\eeq
and thus the eigenvectors are given by
\beq
\vec{\Psi}_{\alpha}=\mathbf{D}\vec{\psi}_{\alpha} ,
\vec{\Phi}_{\alpha}=\vec{\psi}_{\alpha} \ ,
\eeq
which are related as given in \eqref{eq:PhiPsi}.

Similarly, MERW allows for expression of all the eigenvalues and eigenvectors
of the stochastic matrix $\mathbf{P}$ \eqref{eq:Pmerw} in terms of eigenvalues $\lambda_\alpha$
and eigenvectors of $\vec{\psi}_\alpha$ of the adjacency matrix
$\mathbf{A}$
\beq
\Lambda_\alpha = \frac{\lambda_\alpha}{\lambda_0} \ , \ 
\vec{\Psi}_{\alpha} = \mathbf{D}^{1/2}\vec{\psi}_{\alpha} \ , \
\vec{\Phi}_{\alpha} = \mathbf{D}^{-1/2}\vec{\psi}_{\alpha} \ ,
\eeq
where $\mathbf{D}=\uni{diag}(\psi_{01}^2,\psi_{02}^2,\ldots,\psi_{0N}^2)^{-1}$.
In particular, $\Lambda_0=1, \Psi_{0 i}=1, \uni{and}\ \Phi_{0 i}=\psi_{0 i}^2=\pi_{0 i} \uni{\ for \ all}\ i$.
The spectral decomposition of $\mathbf{P}$ then reads
\beq
P_{ij} = \sum_\alpha \Lambda_\alpha \Psi_{\alpha i} \Phi_{\alpha j} =
\sum_\alpha \frac{\lambda_\alpha}{\lambda_0} \psi_{\alpha i}\psi_{\alpha j} \frac{\psi_{0 j}}{\psi_{0 i}} \ .
\label{eq:sd}
\eeq
Thus, clearly all properties of MERW are encoded in the spectral decomposition of the adjacency matrix of a given graph;
it allows for an easier derivation of, for example, the stationary state and dynamical characteristics of MERW for Cayley trees \cite{ZB3,Demo2,Demo1}.

Since methods of both assessing centrality \cite{White} and finding communities \cite{Harel,Latapy}
have utilized calculating powers of the stochastic matrix,
these spectral decompositions allow to make further observations.
The distance matrix used by Latapy and Pons \cite{Latapy} used
\beq
\label{eq:rdist}
r(t)_{ij}=\sqrt \frac{\sum_{k}[(\mathbf{P}^t)_{ik}-(\mathbf{P}^t)_{jk}]^2}{\pi_k} \ ,
\eeq
where $\mathbf{P}$ and $\vec{\pi}$ were meant to correspond to GRW,
and the division by $\pi_k$ was supposed to reduce the effect of a vertex's centrality.
It is mentioned that $\mathbf{r}^2$, the \emph{entrywise} square of this distance matrix, is equivalent to
\beq
\label{eq:rdist_spec}
r^2(t)_{ij}=\sum_{\alpha=1}^{N-1}\Lambda_{\alpha}^{2t}(\Psi_{\alpha i}-\Psi_{\alpha j})^2 \ ,
\eeq
based on spectral decomposition of $\mathbf{P}$ \eqref{eq:Pspec}.

In the case of MERW and GRW (generally, for any RW for which $\mathbf{S}$ defined in \eqref{eq:S} is symmetric),
the spectral decomposition \eqref{eq:sd} leads to the compact form
\beq
\label{eq:rdist_mat}
\mathbf{r}^2(t)=\mathbf{D}[(\mathbf{P}^{2t})_{dg}\mathbf{E}-(\mathbf{P}^{2t})^T]+[\mathbf{E}(\mathbf{P}^{2t})_{dg}-\mathbf{P}^{2t}]\mathbf{D} \ ,
\eeq
where
$(\mathbf{P}^{2t})_{dg}$ is a matrix with $(\mathbf{P}^{2t})_{ii}$ on the diagonal and zeros otherwise.
This is a new formula, which however very much resembles a symmetrized version of a quantity known as \emph{mean first-passage time} matrix.

Mean first-passage time (MFPT) matrix $\mathbf{M}$ is a useful concept for studying RWs. Its elements $M_{if}$ encode the average time to reach the final vertex $f$ from the initial vertex $i$ for the first time.
We invoke a neat construction of the matrix given by Kemeny and Snell \cite{Snell1,Snell2}:
first, we define the \emph{fundamental matrix}
\beq
\label{eq:Z}
\mathbf{Z}=(\mathbf{1}-\mathbf{P}+\vec{e}\vec{\pi}^T)^{-1} \ ,
\eeq
where $\mathbf{1}$ is the identity matrix, and $\vec{\mathbf{e}}=(1,1,...,1)^T$. 
The MFPT matrix is then given by
\beq
\label{eq:MFPT}
\mathbf{M}=(\mathbf{E}\mathbf{Z}_{dg}-\mathbf{Z})\mathbf{D} \ ,
\eeq
where $\mathbf{E}$ is a matrix of all ones, $\mathbf{Z}_{dg}$ is a diagonal matrix with elements $(\mathbf{Z}_{dg})_{ii}=Z_{ii}$, and $\mathbf{D}$ was introduced in \eqref{eq:S}.

The fundamental matrix is defined so as to contain all the powers of the stochastic matrix $\mathbf{P}$, which follows from expansion of $(\mathbf{1}-\mathbf{P})^{-1}$ in a series $\mathbf{1}+\mathbf{P}+\mathbf{P}^2+\ldots$. However, matrix $\mathbf{1}-\mathbf{P}$ is non-invertible and consequently the expansion does not exist. The correction $\vec{e}\vec{\pi}^T$ allows for a well-defined inversion. In fact, instead of the fundamental matrix one may use other so called generalized inverses (the formalism is summarized in \cite{Hunter}), although we use \eqref{eq:Z} for its conceptual and computational simplicity.

In fact, also in \eqref{eq:rdist} we may take $\sum_{t=0}^{\infty} \mathbf{P}^t$ instead of $\mathbf{P}^t$ to account for all the powers of the stochastic matrix, just as above in the case of the MFPT matrix. This infinite sum
\beq
\label{eq:MERWprop}
\pi_f\sum_{t=0}^{\infty}(\mathbf{P}^t)_{fi}=\sqrt{\pi_{f}}G_{fi}\sqrt{\pi_{i}}
\eeq
reproduces the path-integral (MERW) and field-theoretical (GRW) propagator $\mathbf{G}$ of a free relativistic particle, as has been shown in \cite{ZB2}, which supports the view that the stationary probability \eqref{eq:stat} is reminiscent of the square of a wave function.

Nevertheless, the matrix $\mathbf{G}$ needs an elaboration.
Its elementary definition, more general than in \eqref{eq:MERWprop},
is specifically given by the number of paths between any two nodes,
which is just powers of the adjacency matrix $\mathbf{A}^t$ with all path lengths $t$ taken into account.
As the number of paths dramatically grows with their length, however,
a normalizing parameter $e^{\mu}>\lambda_0$ has to be introduced for the sum of paths to converge:
\beq
\label{eq:Gmu}
\mathbf{G}(\mu)=\sum_{t=0}^{\infty} e^{-\mu t} \mathbf{A}^t \ .
\eeq
From the point of view of paths' statistics, $\mathbf{G}(\mu)$ defines the grand-canonical ensemble of paths.
An element $G_{fi}(\mu)$ corresponds to the grand canonical partition function, $\mu$ to the chemical potential,
and the average path length is $\langle t \rangle_{fi} = -(\ln G)'_{fi}(\mu)$.

In equations \eqref{eq:MERWprop} and \eqref{eq:r2}, $\mathbf{G}$ is an abbreviated notation of $\mathbf{G}(\mu_0)$,
where the special choice of $\mu_0=-\ln \lambda_0$ is related to the graph structure by the largest eigenvalue of the adjacency matrix.

While the arbitrary choice of $\mu$ is equivalent to a cut-off of a path length of the order $T=1/\Delta \mu$,
where $\Delta \mu \equiv \mu-\mu_0$,
the limit $\mu \longrightarrow \mu_0$ leads to dominance of infinite paths,
and to a singularity of $\mathbf{G}(\mu)$.
However, $\mu=-\ln \lambda$ very close to $\mu_0$ can be taken, yielding
\beq
\label{eq:Gba}
\mathbf{G}(\mu)=\sum_{t=0}^{\infty}\frac{\mathbf{A}^t}{\lambda^t}=\frac{1}{\lambda}(\lambda\mathbf{1}-\mathbf{A})^{-1} \ ,
\eeq
which is the resolvent of the adjacency matrix.
To define $G(\mu)$ exactly at the singularity $\mu=\mu_0$, the matrix $\lambda_0\mathbf{1}-\mathbf{A}$ has to be projected to the subspace 
perpendicular to $\vec{\psi_0}$ before inversion.
It can be done similarly as in \eqref{eq:Z} by taking
$\left(\lambda_0\mathbf{1}-\mathbf{A}+\lambda_0\vec{\psi_0}\vec{\psi_0}^{T}\right)^{-1}$,
which eliminates the zeroth eigenmode. This is expected and advantageous in community finding methods, as discussed in \cite{ZB4}.

Finally, we obtain
\beq
\label{eq:r2}
\mathbf{r}^2=\mathbf{D}(\mathbf{G}^{2})_{dg}\mathbf{E}-2\sqrt{\mathbf{D}}\mathbf{G}^2\sqrt{\mathbf{D}}+\mathbf{E}(\mathbf{G}^{2})_{dg}\mathbf{D} \ ,
\eeq
with $\mathbf{G}$ functioning as an analogue of the fundamental matrix $\mathbf{Z}$, and where the time dependence has been eliminated.
Thanks to the symmetry of the matrix, however, the singularity of $\mathbf{G}$ cancels out even without the projection discussed in the paragraph above,
in contrast to the definition of MFPT with the use of the fundamental matrix. 


\section{Centrality measures}
\label{sec:centrality}

The above considerations constitute a common framework for a number of centrality measures.
Below, the connections between them are reviewed and established.


\subsection{Centrality based on paths}
\label{sec:paths}
The original concept of counting paths to assess centrality was introduced in 1953 \cite{Katz}. The idea is to count all the paths that lead to a vertex whose importance we measure \cite{Bonacich2}, where the number of paths of length $t$ between vertices $i$ and $f$ of a graph is given by the element $(A^t)_{fi}$ of the $t$-th power of the adjacency matrix. This corresponds exactly to the definition shown in \eqref{eq:Gmu}.

The importance of the final vertex $f$ is then given by the element $I_f$
of the vector $\vec{\mathbf{I}}=(\mathbf{G}(\mu)-\mathbf{1})\vec{\mathbf{e}}\propto \vec{\psi}_0$,
where the uniform vector $\vec{\mathbf{e}}$ was chosen as a set of initial weights,
and the proportionality to the principal eigenvector holds near $\mu_0$.
This special choice thus connects the statistics of paths, a random walk, and the graph structure at the same time.
In the limit $\mu \longrightarrow \mu_0$,
whose nuances are explained at the end of the previous section, the proportionality constant diverges,
and the contribution of other eigenvectors to the centrality is negligible.
In Sec. \ref{sec:comparison}, instead of restricting $\mu$, finite sums are taken,
with the maximal length of the enumerated paths equal to the diameter of a given graph.
The elements of $\vec{\mathbf{I}}$ are squared so that they correspond to the stationary probability of MERW.


The path weights exponential with respect to the path's length were also employed in \cite{White},
although several restrictions on the paths' sets were proposed,
e.g., only the shortest paths, K-short paths or K-short vertex-disjoint paths.
For comparison in Sec. \ref{sec:comparison} we use only the shortest paths without any constraints on their length.

Alternatively, instead of $e^{-\mu t}$, factorial weights $\beta^t/t!$ might be introduced \cite{Estrada1,Estrada2}
(these papers deal primarily with the problem of community finding),
where $\beta$ is a temperature-like parameter tuning the length of paths one wants to account for.
This yields another quantity that may be considered a similarity matrix: the heat kernel $K(-\beta)=(e^{\beta\mathbf{A}})$,
related to heat diffusion or a continuum-time quantum walk, and can be understood as the Green's function of a network of springs.


It is, however, the former weight choice \eqref{eq:Gba} that generates the unique maximally-entropic random walk.
Those weights make MERW directly reflect the structure of the graph,
which is explicit in the transition matrix definition \eqref{eq:Pmerw},
or conversely, appropriately weighted paths gain the interpretation of a random walk.
 

\subsection{Centrality based on powers of the transition matrix}
Equation \eqref{eq:MERWprop} shows that path enumeration is equivalent to the propagation of MERW.
Let us note, however, that the walks are also weighted
by the ratio $\psi_{0f}/\psi_{0i}=\sqrt{\pi_f/\pi_i}$ of stationary probabilities of the two vertices.
It is a reasonable intuition that the importance of a random walk trajectory depends on the importance of the initial and final vertices.
It seems that the problem of calculating centrality by employing the transition matrix
becomes self-consistent (importance calculated from paths, whose weights depend on the importance),
thus, gets rid of arbitrariness.

The method of assessing centrality by
summing consecutive powers of the transition matrix is stated in \cite{White}:
\beq
\label{eq:powers}
\vec{\mathbf{I}}^T=\sum_{t=1}^{T}\vec{\mathbf{\pi}}(0)^T\mathbf{P}^t \ ,
\eeq
where for simplicity we choose uniform initial probability distribution $\vec{\mathbf{\pi}}(0)^T$.
Intuitively, the influence of the initial vertex on its surroundings is estimated with $T$ steps of a random walk.
This parameter controls whether local effects or the stationary state is favored,
with $\vec{\mathbf{I}}$ approaching the stationary probability distribution for large $T$.
The number of steps is usually kept rather small, due to computation costs.

As noted in the previous section, for MERW the definition \eqref{eq:powers} is very similar to counting paths.
Even for relatively small $T$, it produces results very close to the stationary probability distribution.
This is expected for MERW, since its exponential relaxation to the stationary state is on average faster than GRW's
for the benchmark graphs used.
To much extent we have accounted for that behavior, as MERW seems to relax very fast within connected regions (proven for Cayley trees \cite{ZB3,Demo2}),
although it takes long time to relax between two identical connected regions \cite{JO}.


\subsection{Centrality based on mean first-passage times, stationary distributions, and the principal eigenvector of the adjacency matrix}

\begin{figure}[hbtp]
	\centering
		\includegraphics[width=0.5\textwidth]{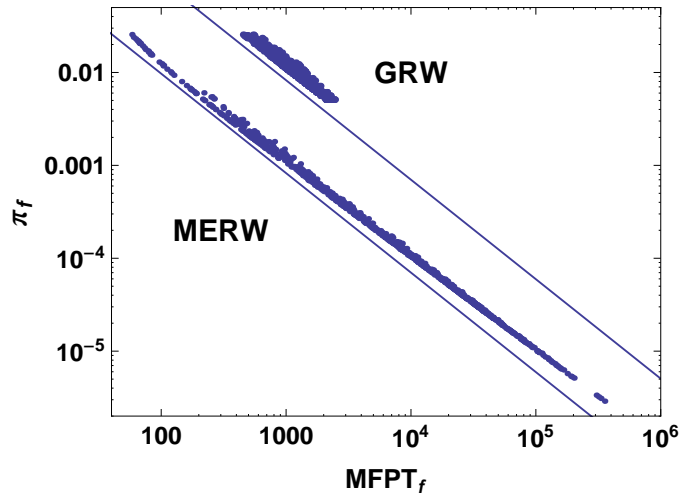}
	\caption{\label{fig:MFPT}
	The stationary probability distribution of MERW and GRW as a function of the averaged rows of MFPT matrix (MFPT of hitting a vertex $f$ averaged over all initial vertices) for a sample graph with $N=1000$ vertices.
	Solid lines have best fit slopes $-1.027\pm0.001$ (MERW) and $-1.070\pm0.005$ (GRW). The correlation for GRW is weaker as the degrees of the graph take values $10-50$ and accordingly the stationary state is quantized.
	$\pi_f$ for GRW is multiplied by 10 for clarity.}
\end{figure}

As shown in Sec. \ref{sec:theory} there is a close analogy between $\mathbf{r}^2$, which uses powers of the transition matrix discussed above,
and mean first-passage times matrix $\mathbf{M}$, introduced in \eqref{eq:MFPT}.
The centrality based on MFPT matrix is given by the inverse of $\sum_{i}M_{if}$,
where the sum represents the average time the information needs to reach the final vertex $f$ from anywhere in the graph.
This definition is called \emph{Markov centrality} in \cite{White}, although the Markov process assumed there is GRW.
For MERW, as demonstrated in Fig.\ref{fig:MFPT}, 
the information extracted from MFPT matrix and the stationary distribution is largely equivalent.

Clearly, the multiplication by $\mathbf{D}$ causes the general trend $M_{if} \sim \pi_{f}^{-1}$.
Since the stationary state of MERW is typically distributed over a wide range of values (even on almost regular graphs \cite{ZB2}), MFPTs correlate with it very strongly and extend much further, especially on bounded-degree graphs, than for GRW, whose stationary distribution is proportional to vertex degrees.
In Fig.\ref{fig:MFPT} the dependence is compared between MERW and GRW.

This observation begs the question: the stationary distribution of which random walk should be chosen to define a centrality measure? Indeed, the one used most widely is GRW, whose stationary state is produced by the simplest version of the prominent PageRank \cite{Page}.
The two random walk centralities have already been compared in \cite{Delvenne} and the conclusion was among others that MERW has "a larger discriminating power between the best and worst pages," and is sensitive to link farms.


However, the connection to other methods has been missing.
Both paths' statistics and random walks are linked to the
idea of calculating centrality as an eigenvector associated with the largest eigenvalue of the adjacency matrix,
which is a concept as old as the economic and sociological papers from 1965 \cite{Hubbell} and 1972 \cite{Bonacich1}.
In the latter, 
this centrality was derived from the assumption that $\vec{\mathbf{I}}_t=\mathbf{A}^t\vec{\mathbf{e}}/\lambda_0^t$
is the $t$-th order popularity measure, and that an objective measure should be taken in $t=\infty$, convergence thus requiring	 the factor $\lambda_0^{-1}$.
Clearly, this formula is simply the canonical ensemble version of the one based on paths \eqref{eq:Gmu},
and the proposed eigenvector centrality is the square root of the stationary state of MERW \eqref{eq:stat}.
This is also reminiscent of the HITS algorithm \cite{HITS}, which nevertheless for directed graphs uses eigenvectors of $\mathbf{A}^T\mathbf{A}$.

We note that just as centrality may be defined with the use of the principal eigenvector of the adjacency matrix or the stochastic matrix (the stationary state vector), there is a family of community detection methods analyzing the rest of the eigenvectors (often it is the spectrum of Laplacian that is analyzed) 
In fact, each of the methods of assessing centrality mentioned above has a number of counterparts that in a similar manner try to find the community structure of a network.
In \cite{ZB4}, a comparison is made between GRW and MERW in performance of some community finding methods based on the concepts presented above.


\section{Comparison}
\label{sec:comparison}

\begin{figure}[htbp]
	\centering
		\includegraphics[width=0.47\textwidth]{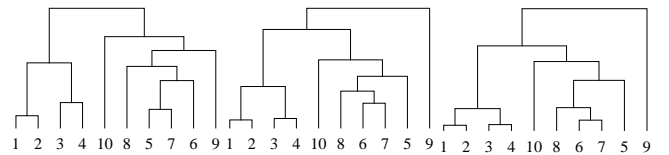}
	\caption{\label{fig:dendrograms}
	The dendrograms correspond to benchmark graphs with the mixing parameter $\mu=0.1,0.3,0.6$ (from left to right).
	1, 2, and 3 represent MERW's stationary state, MFPT, and $\mathbf{P}^t$.
	6, 7, and 8 are GRW's analogues.
	4 denotes weighted paths'
	, and 5 shortest paths' centrality.
	9, 10 are closeness and betweenness.
	In 3 and 8, the maximal power of $\mathbf{P}$ is $T=5$. In 4 and 5 $\mu=\ln \lambda_0$ and $T$ equals the diameter of the graph.
	}
\end{figure}

We check the affinity of different centrality measures described above
(together with the closeness and betweenness centrality given for reference; see \cite{Review})
by comparing the result they produce for a sample of graphs.
For a given graph, each centrality measure produces a vector $\vec{\mathbf{I}}$,
whose consecutive elements are centrality values of the corresponding nodes.
To compare results of a pair of methods on that graph
we take the corresponding pair of vectors
and we measure the root mean square distance between them (cosine or Pearson correlation distance have been checked as well, and have generated similar results).
After repeating this computation for all pairs of centrality measures
we obtain a square matrix.
Since each graph from the sample produces one such matrix,
we take the average (entrywise) over the whole sample.
The entries of the resultant matrix represent the average distance between a pair of centrality measures.
Finally, this distance matrix is used as input for an agglomerative clustering algorithm with average weights,
which generates the dendrograms in Fig.\ref{fig:dendrograms}.
The heights of their branches correspond to the distances between pairs of clusters.
The maximum standard deviation of the distance matrix entries is smaller than $0.61\%$,
hence the results of the clustering algorithm should be correct for most graphs in the sample.

For example, in the dendrogram on the left in Fig.\ref{fig:dendrograms},
the nearest centralities are 1 and 2 (which denote the stationary state of MERW and its MFPT centrality),
and they were the first to be clustered together.
Next, methods 5 and 7 were clustered (the shortest paths' centrality and the MFPT centrality of GRW), and so on.
It can be seen that the closeness and betweenness (9 and 10) are always clustered at the very end,
which means they are very distinct from the other methods.
This is expected, as they are based on different concept of importance,
and could for instance assign a high centrality score to a node near a bottle-neck, even though it is poorly connected.
The methods 4 and 5 depend on the maximum path length $T$ taken into account (it is set to the diameter of the graph, which varies between 4-10),
so their assignment might be different for parameter choices other than shown here.

More importantly, another look at the dendrograms reveals that when the parameters of the benchmark graphs change
there are two groups of methods that do not mix with each other.
One includes centralities derived from MERW (1, 2, and 3: centralities based on the stationary state, MFPT, and $\mathbf{P}^t$, respectively) and 4 which is based on weighted paths,
while the other includes centralities derived from GRW (6, 7, and 8, as for MERW) and 5 which is based on shortest paths.
Thus, methods utilizing GRW are close to each other,
however for graphs with easily distinguishable communities they can cluster together with other centrality measures.
The methods utilizing MERW are connected to path enumeration, as predicted in Sec. \ref{sec:paths},
and they never intermingle with the other centrality measures.
The average distance of this whole group from other methods analyzed is greater than the analogous distance for the group of GRW methods,
whereas the average distance between the members of this group is smaller than the corresponding value for GRW methods.
This means that indeed the centralities defined by MERW comprise a distinct, close-knit family,
and produce equivalent results.


\subsection{Benchmarks}
\label{sec:benchmarks}

In the analysis in the previous section LFR benchmark graphs \cite{Fortunato1} were utilized.
Since they were designed to benchmark community finding algorithms,
they contain communities with preset size distribution, constructed with the use of the \emph{planted partition model}.
Although the graphs are locally random, they model a range of possible real world structures,
and so they serve our purpose in testing centrality measures.

We follow the notation used by the authors of the benchmarks. Thus, by $\mu$ we denote the mixing parameter (this should not be confused with the usage of $\mu$ in \eqref{eq:Gmu}, where chemical potential is meant; context should make it unambiguous, which meaning is intended),
which is the fraction of links a given node shares with the nodes outside its community.
The parameter is approximately equal for all nodes in a graph.
For chosen values of $\mu$, we take $100$ graphs with $N=200$ vertices;
their exponents for the degree and community size distributions are respectively $\tau_1=-2$ and $\tau_2=-1$.
The average and maximum degrees are $10$ and $30$,
and the community sizes range $5-35$.

\section{Conclusions}

In this paper it has been shown that the random walk distance matrix $\mathbf{r}^2(t)$ defined in \eqref{eq:rdist},
when modified 
to account for walks of all lengths,
is equivalent to a symmetric version of the mean-first passage matrix $\mathbf{M}$ \eqref{eq:MFPT},
where the fundamental matrix $\mathbf{Z}$ is substituted with the propagator $\mathbf{G}$.

This observation also leads to the conclusion that a number of known centrality measures are nearly identical
if the random walk under consideration is the maximal-entropy random walk.
This common perspective includes measures related to the properties of graphs:
the eigenvector centrality and centrality based on enumeration of weighted paths,
and those related to random walks: their stationary state, powers of their transition matrix, and finally their MFPT matrix,
as reviewed in Sec. \ref{sec:centrality}.

A numerical investigation on a set of benchmark graphs confirms this thesis,
showing that there is a group of centrality measures related to GRW that tend to produce similar results,
and an even more homogeneous and distinct group of centralities related to MERW.
To quote Bonacich \cite{Bonacich1}:"Three different approaches to calculating popularity scores have almost the same solution [...]. This is an economy; three approaches are reduced to just one. This is the main point of the paper".

\section{Acknowledgements}
Special thanks for numerous discussions are due to Zdzis\l aw Burda. 
Project operated within the Foundation for Polish Science International Ph.D. Projects Programme co-financed by the European Regional Development Fund covering, under the agreement no. MPD/2009/6, the Jagiellonian University International Ph.D. Studies in Physics of Complex Systems.


\begin{thebibliography}{10}

\bibitem{Freeman}
L. Freeman, { Social Networks} \textbf{1}, 215-239 (1979).

\bibitem{WF}
S. Wasserman, and K. Faust, Social Network Analysis: Methods and Applications (Cambridge: Cambridge University Press, 1994).


\bibitem{HITS}
J. Kleinberg, J. Assoc. Comput. Mach. 46, 604 (1999), DOI: http://dx.doi.org/10.1145/324133.324140.

\bibitem{Page}
S. Brin and L. Page, Comput. Networks ISDN Syst. 30, 107 (1998), DOI: http://dx.doi.org/10.1016/S0169-7552(98)00110-X.

\bibitem{Fortunato}
S. Fortunato, { Physics Reports} \textbf{486}, 75-174 (2010).

\bibitem{ZB1}
Z.~Burda, J.~Duda, J.M.~Luck, and B.~Waclaw, Phys. Rev. Lett. \textbf{102}, 160602 (2009).

\bibitem{ZB2}
Z.~Burda, J.~Duda, J.M.~Luck, and B.~Waclaw, { Acta Phys. Pol. B} \textbf{41}, 949 (2010).

\bibitem{BW}
B. Waclaw, {\em Generic Random Walk and Maximal Entropy Random Walk}, Wolfram Demonstration Project.

\bibitem{ZB3}
J.K.~Ochab, and Z.~Burda { Phys. Rev. E} \textbf{85}, 021145 (2012).

\bibitem{Demo2} J.K.~Ochab, 
{\em Dynamics of Maximal Entropy Random Walk and Generic Random Walk on Cayley trees}, Wolfram Demonstration Project.

\bibitem{JO}
J.K.~Ochab, { Acta Phys. Pol. B} \textbf{43} 1143 (2012).

\bibitem{H} 
J.H. Hetherington, Phys. Rev. A \textbf{30}, 2713 (1984).

\bibitem{Evolutionary1}
L.~Demetrius, V.M.~Gundlach and G.~Ochs, 
Theor. Popul. Biol.  \textbf{65}, 211 (2004).

\bibitem{Evolutionary2}
L.~Demetrius, T.~Manke, Physica A \textbf{346}, 682 (2005).

\bibitem{MERW+CN1} 
V. Zlatic, A. Gabrielli, G. Caldarelli, Phys. Rev E \textbf{82}, 066109 (2010). 

\bibitem{Delvenne}
J.-C.~Delvenne, and A.-S.~Libert, { Phys. Rev. E} \textbf{83} 046117 (2011).

\bibitem{MERW+CN3}
R.~Sinatra, J.~Gomez-Gardenes, R.~Lambiotte, V.~Nicosia, and V.~Latora, { Phys. Rev. E} \textbf{83}, 030103 (2011).
\bibitem{MERW+CN4} C. Monthus, T. Garel, 
J. Phys. A: Math. Theor. \textbf{44}, 085001 (2011). 
\bibitem{MERW+CN5} K. Anand, G. Bianconi, S. Severini,
Phys. Rev. E  \textbf{83}, 036109 (2011). 


\bibitem{P}
W. Parry, Trans. Amer. Math. Soc. 112, 55 (1964).

\bibitem{Demo1} J.K.~Ochab, 
{\em Stationary states of Maximal Entropy Random Walk and Generic Random Walk on Cayley trees}, Wolfram Demonstration Project.

\bibitem{White}
S. White, P. Smyth, 
 in:{\em KDD '03: Proceedings of the Ninth ACM SIGKDD International
Conference on Knowledge Discovery and Data Mining, Washington, D.C., 2003}, (ACM, New York, NY, USA, 2003), pp. 266-275.

\bibitem{Harel}
D. Harel, Y. Koren, 
 in:{\em FST TCS '01: Proceedings of the 21st Conference on Foundations of Software Technology
and Theoretical Computer Science,} (Springer-Verlag, London, UK, 2001), pp. 18-41.

\bibitem{Latapy}
M. Latapy, P. Pons, { Lect. Notes Comput. Sci.} \textbf{3733}, 284-293 (2005).

\bibitem{Snell1}
J. G. Kemeny and J. L. Snell, Finite Markov Chains (Springer Verlag, New York, 1976).

\bibitem{Snell2}
C. M. Grinstead and J. L. Snell, Introduction to Probability (American Mathematical Society, Providence, RI, 1997).

\bibitem{Hunter}
J. J. Hunter,
 Res. Lett. Inf. Math. Sci. \textbf{3}, 99-116 (2002).

\bibitem{ZB4}
J.K.~Ochab, and Z.~Burda, Maximal-entropy random walk and community finding,
arXiv:1208.3688v1  [physics.soc-ph].


\bibitem{Katz}
L. Katz, Psychometrika \textbf{18}, 39-43 (1953).

\bibitem{Bonacich2}
P. Bonacich, { Amer. J. Sociol.} \textbf{92 (5)}, 1170-1182 (1987).

\bibitem{Estrada1}
E. Estrada, N. Hatano, { Phys. Rev. E} \textbf{77}(3), 036111 (2008).

\bibitem{Estrada2}
E. Estrada, N. Hatano, { Appl. Math. Comput.} \textbf{214}, 500-511 (2009).


\bibitem{Hubbell}
C. H. Hubbell, Sociometry \textbf{28}, 377-399 (1965).

\bibitem{Bonacich1}
P. Bonacich, { J. Math. Sociol. } \textbf{2 (1)}, 113-120 (1972).

\bibitem{Review}
Dirk Koschützki et al.,
{ Lect. Notes Comput. Sci.} \textbf{3418} 16-61 (2005).

\bibitem{Fortunato1}
A.~Lancichinetti,~S. Fortunato, F.~Radicchi, { Phys. Rev. E} \textbf{78}, 046110 (2008).


\end{thebibliography}
\end{document}